\documentclass[conference]{IEEEtran}

\usepackage{csquotes}
\usepackage{cite}
\usepackage{graphicx}
\graphicspath{ {media/} }
\usepackage{amsmath}
\usepackage{bm}
\usepackage{relsize}
\usepackage{amssymb}
\usepackage{array}
\newcommand{\comm}[1]{}
\usepackage[letterpaper, left=1in, right=1in, bottom=1in, top=0.75in]{geometry}

\begin{document}

\title{RSSI-Based Self-Localization with Perturbed Anchor Positions}

\author{\IEEEauthorblockN{Vikram~Kumar\textsuperscript{*}\textsuperscript{\textdagger}, Reza~Arablouei\textsuperscript{\textdagger}, Raja~Jurdak\textsuperscript{\textdagger}\textsuperscript{*}, Branislav~Kusy\textsuperscript{\textdagger}\textsuperscript{*}, and Neil~W.~Bergmann\textsuperscript{*}}
\IEEEauthorblockA{\textsuperscript{*}School of ITEE, University of Queensland, St Lucia QLD 4072, Australia\\
\textsuperscript{\textdagger}Commonwealth Scientific and Industrial Research Organisation, Pullenvale QLD 4069, Australia}}

\maketitle 

\begin{abstract}
We consider the problem of self-localization by a resource-constrained mobile node given perturbed anchor position information and distance estimates from the anchor nodes. We consider normally-distributed noise in anchor position information. The distance estimates are based on the log-normal shadowing path-loss model for the RSSI measurements. The available solutions to this problem are based on complex and iterative optimization techniques such as semidefinite programming or second-order cone programming, which are not suitable for resource-constrained environments. In this paper, we propose a closed-form weighted least-squares solution. We calculate the weights by taking into account the statistical properties of the perturbations in both RSSI and anchor position information. We also estimate the bias of the proposed solution and subtract it from the proposed solution. We evaluate the performance of the proposed algorithm considering a set of arbitrary network topologies in comparison to an existing algorithm that is based on a similar approach but only accounts for perturbations in the RSSI measurements. We also compare the results with the corresponding Cramer-Rao lower bound. Our experimental evaluation shows that the proposed algorithm can substantially improve the localization performance in terms of both root mean square error and bias. 
\end{abstract}

\begin{IEEEkeywords}
Cooperative localization, multilateration, received signal strength indicator, self-localization, weighted least-squares, wireless sensor networks.
\end{IEEEkeywords}

\section{Introduction}
We consider the problem of self-localization in a network of resource constrained mobile nodes. The mobile node interested in estimating its location is referred to as the \enquote{blind node}. The nodes within the communication range of the blind node, which have their own position information, are referred to as the \enquote{anchor nodes}. An energy-efficient solution to this problem is to assume that the location of the blind node is the same as that of one of the anchor nodes \cite{Jurdak2012,lee2012comon,cluster2016}. In this case, the localization error is a function of  communication range and the anchor position  accuracy. However, for a large communication range, the localization error can be very high. Another solution is to use a multilateration technique. In multilateration, the distance between the blind node and an anchor node is often estimated using received signal strength indicator (RSSI). Then, the position of the blind node is determined as the intersection of all  circles  whose centres are located at the anchor node position coordinates and  radii are equal to the distances of the anchor nodes to the blind node. Given accurate distances and anchor positions, the intersection of the circles can be computed using the least-squares method. In practice, the RSSI measurements are subject to perturbations caused by model inaccuracy, thermal noise, measurement error, etc. The anchor position information is also  corrupted by noise or error, especially when it is the product of a previous estimation process including the global positioning system (GPS). 

Two main classes of approaches  deal with the adverse effects  of  RSSI and anchor position perturbations on localization performance. The first class is based on the solution of a linearized system of equations using  weighted least-squares (WLS), total least-squares (TLS), element-wise total least-squares (EWTLS), etc.  \cite{chan1994simple,ho2012bias,noroozi2016weighted,wan2009localization}. The second approach is based on   optimization techniques such as  semidefinite programming (SDP) or second-order cone programming (SOCP) \cite{Mekonnen2014, Naddafzadeh-Shirazi2014,  Angjelichinoski2014}. The SDP or SOCP-based solutions have better performance over TLS-based solutions but incur high computational cost. Most of the TLS-based solutions either assume perfect knowledge of the anchor positions or assume normal distribution for errors in both RSSI-based distance measurements and anchor positions. In practice, the perturbations of the distance estimates from RSSI measurements in the widely-accepted log-normal path-loss shadowing model do not have normal distribution. The overdetermined system of linear equations representation $\tilde{\mathbf{A}}\textbf{w} \approx \tilde{\textbf{b}} $ of the considered localization problem has the  components  of both distance and anchor position perturbations in $\tilde{\textbf{b}}$ (details given in section IV). The presence of the distance perturbations in $\tilde{\textbf{b}}$ leads to an error distribution different than $\tilde{\mathbf{A}}$. This violates the basic assumption of identical error distribution in TLS and EWTLS techniques \cite{markovsky2007overview, premoli2002parametric}. 

In this paper, we propose a closed-form  solution to localize a blind node when perturbations are present in both RSSI and anchor position information. We assume normal distribution for the perturbations in the anchor position information. We consider shadowing path-loss model for the RSSI measurements with normally-distributed perturbation in the logarithmic domain leading to log-normal distribution for  the distance estimates. Our algorithm produces a bias-compensated weighted least-squares  solution.  The weights  are estimated by taking into account the statistical properties of perturbations in both RSSI and anchor position information. In addition, we calculate and compensate for the bias in the solution caused by perturbed anchor position and RSSI measurements. We evaluate our algorithm using numerical simulations with various  arrangements of the blind node and the anchor nodes within arbitrary network topologies. Due to similar computational requirements, the performance of the proposed algorithm is compared with an existing WLS-based hyperbolic algorithm \cite{Tarrio2011} that only takes into account the perturbations in the RSSI measurements. The simulation results show that the proposed algorithm offers significant improvement in the localization performance over the hyperbolic algorithm of \cite{Tarrio2011}.

This work is useful in emerging cooperative wireless networks of mobile nodes such as environmental monitoring and long-term tracking of equipment or wildlife  in large outdoor areas \cite{Jurdak2012, Atts2015, Allen2015, ahmad2015experiments}. Typically, such networks are very large and contain only few  nodes with independent positioning capabilities such as GPS. The rest of nodes in the network depend upon cooperative positioning techniques to obtain a position estimate. The  poor performance of the GPS in indoor, dense forests, urban areas, or even with a cloudy sky results in inaccuracies/uncertainties in the form of perturbations in the anchor position information \cite{Vallina-Rodriguez2013}.

\section{Related Work}
In \cite{Tarrio2011}, the authors propose two WLS algorithms, called hyperbolic and circular, to localize a node using RSSI-based distance measurements while assuming the log-normal shadowing path-loss model. The hyperbolic algorithm linearizes the problem and solves it using the WLS method. The circular algorithm  minimizes the weighted approximation of the original non-linear sum-square-error cost function using the gradient-descent method. The circular algorithm performs better than the hyperbolic algorithm due to minimization of the original cost function by an iterative approach. The proposed hyperbolic algorithm match the low computational requirement of our applications of interest but do not consider perturbations in the anchor node positions. Our proposed algorithm improves the hyperbolic algorithm by considering perturbations in the anchor position information as well.
The authors of \cite{shi2014distributed} propose an element-wise weighted total least-squares  solution to localize a node in the presence of errors in both  anchor position and distance measurements.  The EWTLS \cite{markovsky2006element},\cite{premoli2002parametric} solutions deal with the element-wise variance but assume the same type of error distribution. In case of the RSSI-based localization, the distance measurements follow log-normal distribution whereas anchor position information is corrupted with normally-distributed noise. In \cite{li2016accurate}, a closed-form solution for node localization using noisy TDOA measurements and anchor position information is presented. The authors assume the scenario of large equal radius and develop the estimator based on a geometric approach removing the need for introducing an auxiliary variable. The authors of \cite{weng2011total} propose a TLS based solution for location estimation of stationary source  using noisy TDOA measurements and perturbed anchor positions. 
In \cite{wan2009localization}, the authors present three different  solutions based on Taylor series, WLS and constrained TLS (CTLS) for localization in the presence of additive zero-mean Gaussian noise in both distance and anchor position information. The Taylor series and WLS-based solutions are proposed for the scenario where perturbations are only present in distance measurements. In the CTLS-based algorithm, the authors assume perturbation in anchor position information as well. The authors of \cite{zhou2012multilateration} also consider the presence of independent additive zero-mean Gaussian noise in both distance and anchor position information. They present another CTLS-based localization algorithm for the case where distance measurement errors are negligible  compared to anchor position errors.

The authors in \cite{noroozi2016weighted},\cite{noroozi2017target}, propose a WLS-based closed-form solution to determine the position of a mobile target in the presence of multiple transmitters and receivers using TDOA measurements is proposed. The algorithm is based on the intersection of the ellipsoids defined by bistatic range measurements from a number of transmitters and receivers. These works are based on the TDOA measurements and assume the perfect knowledge of the anchor positions. However, in this paper, we consider the  problem of RSSI-based  self-localization in the presence of perturbations in both  RSSI and anchor position information. 

In summary, most of the previous works assume perfect knowledge of anchor positions or additive zero-mean Gaussian perturbation in distance estimate as well. From an algebraic point of view, in a system of linear equation representation $\tilde{\mathbf{A}}\textbf{w} = \tilde{\textbf{b}} $ of the considered localization problem, perturbation in distance estimates  only affects the elements of $\textbf{b}$. The LS and WLS techniques in general, are designed to take such errors into account. The works considering error in both anchor position and distance measurement assume a similar type of error distribution and present solutions based on TLS, EWTLS or CTLS. TLS and EWTLS techniques are able to take into account errors in both $\mathbf{A}$ and $\mathbf{b}$ . They require that the noise components in both $\tilde{\mathbf{A}}$ and $\tilde{\mathbf{b}}$ to be  independent and identically distributed. In the considered multilateration localization, however, it can be shown that the noise components in the formulated equations are algebraically related. The CTLS-based technique \cite{abatzoglou1987constrained} can account for the linear correlation of errors in $\mathbf{A}$ and $\mathbf{b}$  but, in the considered multilateration localization problem, the dependence is not linear. 

In contrast to the existing works, we consider more realistic perturbation scenario  for the RSSI-based distance measurements.We consider shadowing path-loss model for the RSSI measurements with normally-distributed perturbation in the logarithmic domain leading to log-normal distribution for  the distance estimates. We assume normal distribution for the perturbations in the anchor position information. We present a solution using WLS where weights are calculated considering the statistical properties of noise in anchor position and RSSI measurements. In addition, we  derive and compensate for the bias in the proposed solution. 

\section{Problem Statement}
Consider a mobile node, referred to as the blind node, that is unaware of its true position $(x_b,y_b)$ and is interested in self-localization using noisy measurements of positions and distances  from nodes within its communication range, referred to as the anchor nodes. Although we assume two dimensional coordinates in this paper, the extension to higher dimension is straight forward. Let there be $M\geq3$ arbitrarily dispersed anchor nodes   at locations $(x_i,y_i)$, $i=1,...,M$. The noisy anchor position information at blind node is denoted by $(\tilde{x}_{i},\tilde{y}_{i})$, $i=1,...,M$. The blind node calculates its distance from anchor nodes using the RSSI measurements denoted by $\tilde{p_{i}}$, $i=1,...,M$. We make the following common assumptions regarding the perturbations:
 
\textit{A1}: The  perturbations of anchor position  are additive independent zero-mean Gaussian  with known standard deviation. The standard deviation $\sigma_{a_{i}}$ in the $ i$th anchor node is the same for both x and y axes. The perturbations of positions of different anchor nodes  are independent of each other and may have different value of $\sigma_{a_{i}} $. Therefore, we have

\begin{equation*}
\tilde{x}_i = x_{i} + {n}_{x_i} ,~   \tilde{y}_i = y_{i} + {n}_{y_i}
\end{equation*} 
\begin{equation*}
{n}_{x_i}, {n}_{y_i} \sim \mathcal{N}(0,\sigma_{a_{i}}).
\end{equation*}    

\textit{A2}: The considered radio signal path-loss model is  log-normal shadowing. Therefore, the RSSI measurement at the blind node for the signal transmitted from the $i$th anchor node, denoted by $\tilde{p}_{i\text{(dBm)}}$ in the logarithmic (dBm) scale, has a Gaussian distribution with  mean $\bar{p}_{i\text{(dBm)}}$ and standard deviation $\sigma_{p_{i}}$, i.e.,
\begin{equation}\label{pidbm1}
\tilde{p}_{i\text{(dBm)}} = \bar{p}_{i\text{(dBm)}} + n_{\sigma_{p_{i}}}
\end{equation}
\begin{equation*}
n_{\sigma_{p_{i}}}  \sim \mathcal{N}(0,\sigma_{p_{i}}).
\end{equation*}  

The shadowing path-loss model describes the relationship between the $i$th mean power and the distance between the blind node and the $i$th anchor node, i.e.,
\begin{equation*}
d_i=\sqrt{\left(x_i-x_b\right)^2+\left(y_i-y_b\right)^2},
\end{equation*}
as
\begin{equation}\label{pidbm2}
\bar{p}_{i\text{(dBm)}} = p_{0\text{(dBm)}} - 10\eta\log_{10}{\frac{d_i}{d_0}}
\end{equation}
where $d_0$, $p_{0\text{(dBm)}}$, and $\eta$, are the reference distance, the received power at the reference distance, and the path loss exponent, respectively. Therefore, given the perturbed value $\tilde{p}_{i\text{(dBm)}}$, the RSSI-induced estimate for the distance between the blind node and the $i$th anchor node, denoted by $\tilde{d}_{i}$, is calculated as 
\begin{equation}\label{d_i_p_i_relation}
\tilde{d}_i = d_0 10^{\dfrac{\tilde{p}_{i\text{(dBm)}} - p_{0\text{(dBm)}}}{10\eta}}.
\end{equation}
Additionally, we assume that the blind node and the anchor nodes have limited computational and energy resources. Hence, at any particular instance of localization, only one RSSI measurement and position estimate from each anchor node is available to the blind node. 
\section{Proposed Algorithm}
Given anchor positions  $\left(x_i,y_i \right)$, $i = 1,...,M $, and the corresponding distances $d_{i}$,~$i = 1,...,M $,  the blind node position can be calculated as the intersection of the circles defined by 
\begin{equation}\label{org_circle}
\left(x - x_i\right)^2 + \left(y - y_i\right)^2 = d_i^2, i = 1,...,M.
\end{equation}
Subtracting the equation for $ i = 1$ from the others results in a system of linear equations expressed as 
\begin{equation}\label{org_mat}
2\mathbf{A}\textbf{w}= \textbf{b}
\end{equation}
where
\begin{equation*}
\begin{split}
&\mathbf{w} = 
\begin{bmatrix}
x\\y
\end{bmatrix},\\
&\mathbf{A} = \begin{bmatrix}
x_2 - x_1 & y_2 - y_1 \\
x_3 - x_1 & y_3 - y_1 \\
       ... & ... \\
x_i - x_1 &  y_i - y_1  
\end{bmatrix},\\
&\mathbf{b} = \begin{bmatrix}
d_1^2  - d_2^2 + k_2 - k_1 \\
d_1^2  - d_3^2 + k_3 - k_1\\
...\\
d_1^2  - d_i^2 + k_i - k_1
\end{bmatrix},
\end{split}
\end{equation*}
and
\begin{equation*}
k_i = x_i^2 + y_i^2. 
\end{equation*}
The well-known LS solution of \eqref{org_mat} is 
\begin{equation}\label{org_x_sol}
\hat{\textbf{w}} = \frac{1}{2}\left(\mathbf{A}^T \mathbf{A}\right)^{-1}\mathbf{A}^T\mathbf{b}.
\end{equation}
However, we do not have access to the unperturbed values $x_i$, $y_i$, and $d_i$. Hence, we replace them with their corresponding perturbed observations  $\tilde{x}_i$, $\tilde{y}_i$ and $\tilde{d}_i$.

To factor in the difference in the scale and statistical properties of the values associated with different anchor nodes, we use the WLS solution given as
\begin{equation}\label{noisy_mat_sol}
\hat{\mathbf{w}} = \frac{1}{2}\left(\tilde{\mathbf{A}}^T \mathbf{S}^{-1} \tilde{\mathbf{A}}\right)^{-1}\tilde{\mathbf{A}}^T\mathbf{S}^{-1}\tilde{\mathbf{b}}
\end{equation}
where
\begin{equation*}
\begin{split}
&\hat{\mathbf{w}} = 
\begin{bmatrix}
\hat{x}\\\hat{y}
\end{bmatrix},\\
&\tilde{\mathbf{A}} = \begin{bmatrix}
\tilde{x}_2 - \tilde{x}_1 & \tilde{y}_2 - \tilde{y}_1 \\
\tilde{x}_3 - \tilde{x}_1 & \tilde{y}_3 - \tilde{y}_1 \\
       ... & ... \\
\tilde{x}_M - \tilde{x}_1 &  \tilde{y}_M - \tilde{y}_1  
\end{bmatrix},\\
&\tilde{\mathbf{b}} = \begin{bmatrix}
\tilde{d}_1^2  - \tilde{d}_2^2 + \tilde{k}_2 - \tilde{k}_1 \\
\tilde{d}_1^2  - \tilde{d}_3^2 + \tilde{k}_3 - \tilde{k}_1\\
...\\
\tilde{d}_1^2  - \tilde{d}_M^2 + \tilde{k}_M - \tilde{k}_1
\end{bmatrix},
\end{split}
\end{equation*}
and
\begin{equation*}
\tilde{k}_i = \tilde{x}_i^2 + \tilde{y}_i^2. 
\end{equation*}

In \eqref{noisy_mat_sol}, $\mathbf{S}$ is the covariance matrix of $\tilde{\mathbf{b}}$ with its $(i,j)$th entry given by
\begin{equation}\label{S_ij1}
s_{ij} = \begin{cases}
\mathrm{Var}(\tilde{d}_1^2  - \tilde{d}_{i+1}^2 + \tilde{k}_{i+1} - \tilde{k}_1) & \text{if} \ i=j\\
\mathrm{Var}(\tilde{d}_1^2 - \tilde{k}_1) & \text{if} \ i\neq j
\end{cases}
\end{equation}
Considering the assumptions \textit{A1} and \textit{A2} in addition to the independent nature of the perturbations of the anchor positions and the RSSI-induced distances, \eqref{S_ij1} can be written as 
\begin{equation}\label{S_ij2}
\resizebox{\linewidth}{!}{
$s_{ij} = \begin{cases}
\mathrm{Var}(\tilde{k}_{i+1}) + \mathrm{Var}(\tilde{k}_1) + \mathrm{Var}(\tilde{d}_1^2)  + \mathrm{Var}(\tilde{d}_{i+1}^2) & \text{if} \ i=j\\
\mathrm{Var}(\tilde{k}_1)+\mathrm{Var}(\tilde{d}_1^2)   & \text{if} \ i\neq j
\end{cases}$
}\end{equation}

To calculate $\mathrm{Var}(\tilde{k}_{i})$, we note that $\tilde{k}_{i}$ is the sum of squares of independent normally distributed random variables $\tilde{x}_i $  and $\tilde{y}_i $ with non-zero mean. Therefore, $\tilde{k}_{i}/\sigma_{a_{i}}^2$ has a non-central chi-squared distribution  with the variance
\begin{equation}
 \mathrm{Var}\left(\frac{\tilde{k}_{i}}{\sigma_{a_{i}}^2}\right) = 4\left(1 + \frac{x_i^2 + y_i^2}{\sigma_{a_{i}}^2}\right) \end{equation}
and consequently
\begin{equation}\label{Var(k_i)}
 \mathrm{Var}\left(\tilde{k}_{i}\right) = 4\sigma_{a_i}^2\left(\sigma_{a_i}^2 + x_i^2 + y_i^2\right).  
\end{equation}
Considering the assumption \textit{A2}, $\mathrm{Var}({\tilde{d_i^2}})$ is calculated as \cite{Tarrio2011}
\begin{equation}\label{Var(d_tilde)}
\mathrm{Var}({\tilde{d_i^2}})=\mathrm{exp}(4\mu_{d_i})\left[\mathrm{exp}(8\sigma_{d_i}^2)-\mathrm{exp}(4\sigma_{d_i}^2)\right]
\end{equation}
where

\begin{equation*}
\mu_{d_i} = \ln d_i,\ \ 
\sigma_{d_i} = \frac{\ln10}{10\eta}\sigma_{p_{i}}.
\end{equation*}
As $x_i$, $y_i$, and $d_i$ are not available, we replace them with their corresponding perturbed observations $\tilde{x}_i$, $\tilde{y}_i$, and $\tilde{d}_i$ in \eqref{Var(k_i)} and \eqref{Var(d_tilde)}.

We observe that the perturbation effecting $\tilde{\textbf{b}}$ is not neither additive nor zero-mean. Therefore, the solution given by \eqref{noisy_mat_sol} is biased. The expectation of the $i$th entry of $\tilde{\textbf{b}}$, is calculated as  
\begin{equation}\label{Eb_i}
\mathop{\mathbb{E}[\tilde{b}_i]} = \mathop{\mathbb{E}[\tilde{d}_1^2  - \tilde{d}_i^2 + \tilde{k}_i - \tilde{k}_1]}.
\end{equation}
Considering the independent nature of the perturbations of the anchor positions and RSSI-induced distances, \eqref{Eb_i} can be rewritten as 
\begin{equation}\label{b_i1}
\mathop{\mathbb{E}[\tilde{b}_i]} = \mathop{\mathbb{E}[\tilde{d}_1^2]}  - \mathop{\mathbb{E}[\tilde{d}_i^2]} + \mathop{\mathbb{E}[\tilde{k}_i]} - \mathop{\mathbb{E}[\tilde{k}_1]}.
\end{equation}

To calculate $\mathop{\mathbb{E}[\tilde{d}_i^2]}$, we perceive that  $\tilde{d}_i^2$ using \eqref{pidbm1}-\eqref{d_i_p_i_relation} is equal to 
\begin{equation*}
\begin{split}
\tilde{d}_i^2=d_i^2\exp\left({\sqrt{2}un_{p_i}}\right)
\end{split}             
\end{equation*}
where
\begin{equation}
u= \frac{\ln10}{5\sqrt{2}\eta}.
\end{equation}
Thus, we have
\begin{equation}\label{E_di^2}
\begin{split}
\mathop{\mathbb{E}[\tilde{d}_i^2]}
&= d_i^2\mathop{\mathbb{E}\left[\exp\left({\sqrt{2}un_{p_i}}\right)\right]}\\
&= d_i^2\exp\left({u^2\sigma_{n_{p_i}}^2}\right).
\end{split}
\end{equation}
Note that $u^2\sigma_{n_{p_i}}^2$ is small even for high values of $\sigma_{n_{p_i}}$ such as $5$dB. Therefore, using the second-order Taylor-series expansion of the function $\mathrm{exp}(u^2\sigma_{n_{p_i}}^2)$ around zero, \eqref{E_di^2} is approximated as
\begin{equation}\label{E_d_i_F}
\mathop{\mathbb{E}[\tilde{d}_i^2]} = d_i^2 + d_i^2\left(u^2\sigma_{n_{p_i}}^2 + \frac{u^4\sigma_{n_{p_i}}^4}{2}\right).
\end{equation}
The term $\mathop{\mathbb{E}[\tilde{k}_i]}$ in \eqref{b_i1} is equal to 
\begin{equation}\label{E_k_i}
\mathop{\mathbb{E}[\tilde{k}_i]} = x_i^2 + y_i^2 + 2\sigma_{n_{a_i}}^2.
\end{equation}

Using  \eqref{E_d_i_F} and \eqref{E_k_i}, $\mathop{\mathbb{E}[\tilde{\textbf{b}}]}$ can be written as 
\begin{equation}
\mathop{\mathbb{E}\left[\tilde{\textbf{b}}\right]} = \textbf{b} +\textbf{c}
\end{equation}
where the $i$th entry of $\mathbf{c}$ is
\begin{equation*}
c_i = \left(u^2\sigma_{n_{p_i}}^2 + \frac{u^4}{2}\sigma_{n_{p_i}}^4 \right)\left(d_1^2 - d_i^2\right) + 2\left(\sigma_{n_{a_i}}^2 - \sigma_{n_{a_1}}^2\right).
\end{equation*}
Since we do not have access to $d_i $, we use the corresponding perturbed values in the calculation of $\textbf{c}$. Hence, the bias of the solution \eqref{noisy_mat_sol} due to $\mathop{\mathbb{E}[\tilde{\mathbf{b}}]} \neq \textbf{0}$ can be calculated as
\begin{equation}
\mathop{\mathbb{E}[\hat{\mathbf{w}}- \mathbf{w}]} = \frac{1}{2}\left(\tilde{\mathbf{A}}^T \mathbf{S}^{-1} \tilde{\mathbf{A}}\right)^{-1}\tilde{\mathbf{A}}^T\mathbf{S}^{-1}\mathbf{c}.
\end{equation}
Consequently, we express the bias-compensated WLS solution as 
\begin{equation}\label{proposed}
\hat{\mathbf{w}} = \frac{1}{2}\left(\tilde{\mathbf{A}}^T \mathbf{S}^{-1} \tilde{\mathbf{A}}\right)^{-1}\tilde{\mathbf{A}}^T\mathbf{S}^{-1}\left(\tilde{\mathbf{b}} - \textbf{c}\right).
\end{equation}

\section{Experimental Setup}
\begin{figure*}
    \centering   
    \includegraphics[width=\textwidth]{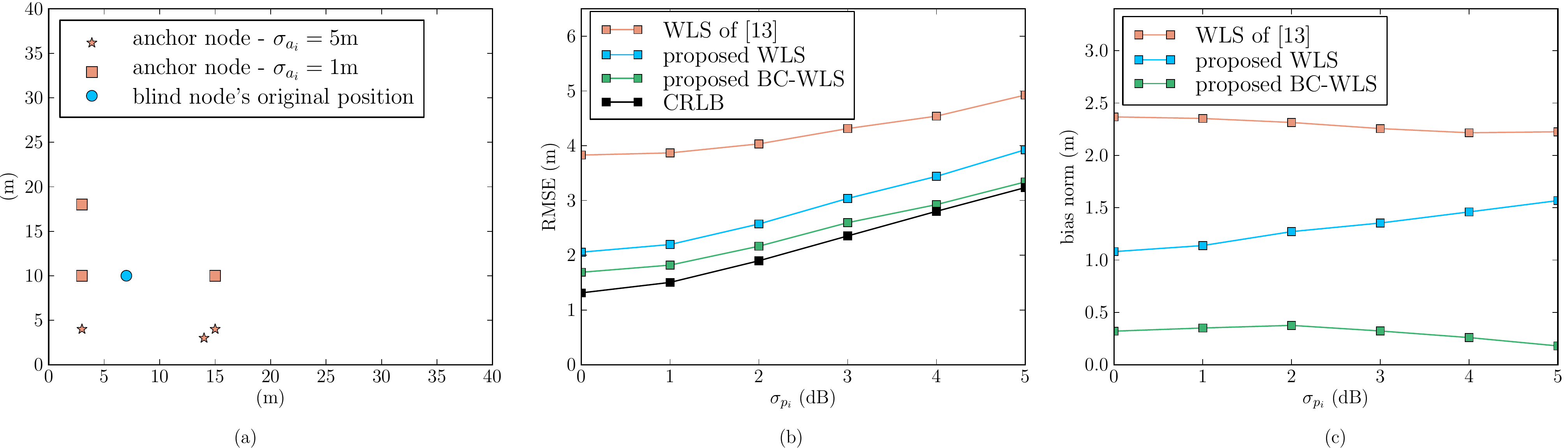}  
    \caption{(a) the network topology; (b) the RMSE \begingroup
\tiny
$\left(\sqrt{\mathop{\mathbb{E}\left[\left\Vert \hat{\mathbf{w}} -\mathbf{w} \right\Vert^2\right]}}\right)$\endgroup~ of the proposed and existing algorithms as well as the corresponding CRLB for different values of $\sigma_{p_i}$; (c) the bias norm  $\left(\left\Vert\mathop{\mathbb{E}\left[ \hat{\mathbf{w}} -\mathbf{w} \right]} \right\Vert_2 \right)$ of the proposed and existing algorithms for different values of $\sigma_{p_i}$.}
    \label{fig:SHTS1}
\end{figure*}  
\begin{figure*} 
    \centering 
    \includegraphics[width=\textwidth]{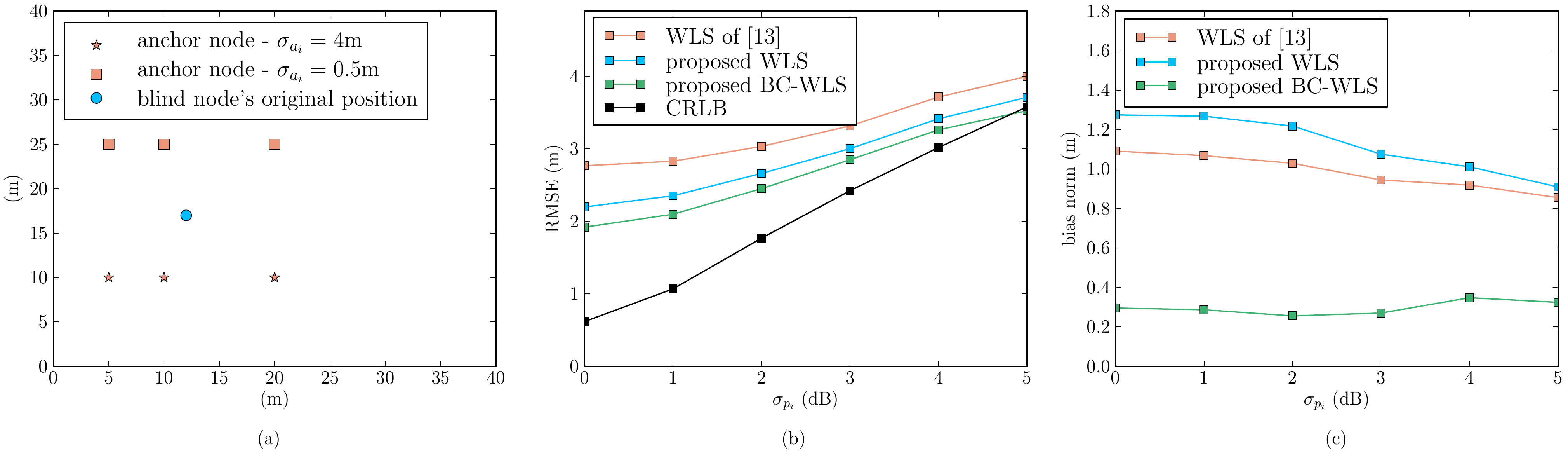}  
    \caption{(a) the network topology; (b) the RMSE \begingroup
\tiny
$\left(\sqrt{\mathop{\mathbb{E}\left[\left\Vert \hat{\mathbf{w}} -\mathbf{w} \right\Vert^2\right]}}\right)$\endgroup~ of the proposed and existing algorithms as well as the corresponding CRLB for different values of $\sigma_{p_i}$; (c) the bias norm $\left(\left\Vert\mathop{\mathbb{E}\left[ \hat{\mathbf{w}} -\mathbf{w} \right]} \right\Vert_2 \right)$ of the proposed and existing algorithms for different values of $\sigma_{p_i}$.}
    \label{fig:SHTS2}    
\end{figure*}
We consider a $40\text{m}\times40\text{m}$ area to evaluate the performance of the proposed algorithm. We assume that the anchor nodes estimate their positions via the GPS in a heterogeneous GPS performance environment. Hence, the  anchor nodes have  different GPS error standard deviations. In practice, if the GPS operational context, e.g., surroundings or hardware, is not the same for all the nodes, then it results in a heterogeneous scenario. As an example, the GPS localization performance is a function of the time during which the GPS receiver is active \cite{jurdak2010adaptive, jurdak2013camazotz}. Therefore, variations in the GPS activity time may lead to a heterogeneous scenario. The heterogeneous scenario may also arise in resource constrained  cooperative or group/cluster-based  localization where the GPS activation time is decided based on the energy budget of the individual nodes. 

The performance of RSSI-based localization highly depends on the network geometry \cite{denkovski2016geometric}. To emulate the geometries encountered by mobile nodes, we evaluate the performance of the proposed algorithm with various arbitrarily selected anchor and blind node positions. We  consider the standard deviation of the GPS noise to be in the range of 0.5m to 5.0m. The RSSI measurement errors range from 0dB to 5dB in all the experiments. The values of the path-loss model parameters used in our experiments are $d_0=1$m, $p_{0(\text{dBm})}=-33.44$ and $n_p =3.567$. These values are based on the results reported in \cite{ahmad2015experiments}.

We compare the performance of the proposed algorithm with that of the so-called weighted hyperbolic algorithm proposed in \cite{Tarrio2011}. This algorithm  produces a WLS solution based on the assumption that only the RSSI measurements are corrupted by noise/error and the anchor node positions are known exactly  at the blind node. Therefore, the proposed algorithm can be viewed as an improvement over the algorithm of \cite{Tarrio2011} taking into account the effects of perturbation in anchor node positions. In addition, we present CRLB on the localization error derived in \cite{GDIEEE}. 

\section{Simulation Results}
We use the root mean square error (RMSE) and bias norm as the performance measures in our evaluations. The results are averaged over 10,000 independent trials. We represent the proposed bias-compensated WLS-based solution as \enquote{BC-WLS}. The results for the solution without bias compensation are labeled as \enquote{proposed WLS}. The hyperbolic algorithm of \cite{Tarrio2011} is marked as \enquote{WLS of \cite{Tarrio2011}}. 

The results for an arbitrarily selected node arrangement where there are three anchor nodes with $\sigma_{a_i}=5$m and three anchor nodes with $\sigma_{a_i}=1$ are given in Fig. \ref{fig:SHTS1}. The proposed BC-WLS algorithm significantly outperform the WLS algorithm of \cite{Tarrio2011} in terms of both RMSE and bias norm. As an example, for low values of $\sigma_{p_i} = 0$dB and 1dB, the RMSE of the proposed BC-WLS algorithm is half of that of the WLS algorithm of \cite{Tarrio2011}. Similarly for $\sigma_{p_i} =5$dB, the BC-WLS algorithm reduces the RMSE to 3m from 5m in WLS of \cite{Tarrio2011}. In addition, the proposed BC-WLS algorithm has a very small bias whereas the bias of WLS algorithm of \cite{Tarrio2011} is around 2.5m. It is to be noted that the RMSE of the proposed BC-WLS algorithm is close to CLRB.

The results for another arbitrarily selected node arrangement is given in Fig. \ref{fig:SHTS2}. In this experiment, there are three anchor nodes with $\sigma_{a_i}=4$m and three anchor nodes with $\sigma_{a_i}=0.5$m. The proposed BC-WLS algorithm performs well as compared to the WLS algorithm of \cite{Tarrio2011} in this node arrangement as well. For all the considered RSSI noise values, the proposed BC-WLS algorithm reduces the RMSE by roughly 1m. In addition, the proposed solution has a relatively small bias. 

In summary, the proposed BC-WLS algorithm significantly outperforms the WLS algorithm of \cite{Tarrio2011} in the realistic scenarios of heterogeneous errors in anchor positions given distance estimates based on perturbed RSSI measurements.

\section{Conclusion}
We considered the problem of self-localization  in the presence of perturbation in the RSSI measurements as well as the anchor positions information. We proposed a closed-form biased-compensated weighted least-squares solution. We evaluated the performance of the proposed algorithm in comparison with a previously-proposed algorithm that only accounts for perturbations in the RSSI measurements considering several arbitrary arrangements of the anchor nodes and the blind node. Our simulation results showed that the proposed algorithm can significantly reduce the localization RMSE and bias.

\bibliographystyle{IEEEtran}

\bibliography{References} 

\end{document}